\documentclass[aps,pra,reprint,superscriptaddress,floatfix,nofootinbib,longbibliography]{revtex4-2}
\usepackage{graphicx}
\usepackage{amssymb}
\usepackage{amsmath}
\usepackage{color}
\usepackage{bbm}
\usepackage{bm}
\usepackage{hyperref}
\usepackage{amsthm}
\usepackage{subfigure}
\usepackage{bbold}
\newtheorem{theorem}{Theorem}
\usepackage{array}

\DeclareMathOperator{\tr}{tr}

\newcommand{\beq}{\begin{eqnarray}}
\newcommand{\eeq}{\end{eqnarray}}

\newcolumntype{M}[1]{>{\centering\arraybackslash}m{#1}}

\begin{document}
\title{Quantum steering as a witness of quantum scrambling}
\author{Jhen-Dong Lin}
\thanks{Jhen-Dong Lin and Wei-Yu Lin contributed equally to this work.}
\affiliation{Department of Physics, National Cheng Kung University, 701 Tainan, Taiwan}
\affiliation{Center for Quantum Frontiers of Research \& Technology, NCKU, 70101 Tainan, Taiwan}

\author{Wei-Yu Lin}
\thanks{Jhen-Dong Lin and Wei-Yu Lin contributed equally to this work.}
\affiliation{Department of Electrical Engineering, National Taiwan University, 106 Taipei, Taiwan}

\author{Huan-Yu Ku}
\affiliation{Department of Physics, National Cheng Kung University, 701 Tainan, Taiwan}
\affiliation{Center for Quantum Frontiers of Research \& Technology, NCKU, 70101 Tainan, Taiwan}

\author{Neill Lambert}
\affiliation{Theoretical Quantum Physics Laboratory, RIKEN Cluster for Pioneering Research, Wako-shi, Saitama 351-0198, Japan}

\author{Yueh-Nan Chen}
\email{yuehnan@mail.ncku.edu.tw}
\affiliation{Department of Physics, National Cheng Kung University, 701 Tainan, Taiwan}
\affiliation{Center for Quantum Frontiers of Research \& Technology, NCKU, 70101 Tainan, Taiwan}

\author{Franco Nori}
\affiliation{Theoretical Quantum Physics Laboratory, RIKEN Cluster for Pioneering Research, Wako-shi, Saitama 351-0198, Japan}
\affiliation{RIKEN Center for Quantum Computing (RQC), Wakoshi, Saitama 351-0198, Japan}
\affiliation{Department of Physics, The University of Michigan, Ann Arbor, 48109-1040 Michigan, USA}

\date{\today}

\begin{abstract}
Quantum information scrambling describes the delocalization of local information to global information in the form of entanglement throughout all possible degrees of freedom. A natural measure of  scrambling is the tripartite mutual information (TMI), which quantifies the amount of delocalized information for a given quantum channel with its state representation, i.e., the Choi state. In this work, we show that quantum information scrambling can also be witnessed by temporal quantum steering for qubit systems. We can do so because there is a fundamental equivalence between the Choi state and the pseudo-density matrix formalism used in temporal quantum correlations. In particular, we propose a quantity as a scrambling witness, based on a measure of temporal steering called temporal steerable weight. We justify the scrambling witness for unitary qubit channels by proving that the quantity vanishes whenever the channel is non-scrambling.
\end{abstract}

\maketitle

\section{Introduction.}
Quantum systems evolving under strongly interacting channels can experience the delocalization of initially local information into non-local degrees of freedom. Such an effect is termed ‘‘quantum information scrambling,'' and this new way of looking at delocalization in quantum theory has found applications in a range of physical effects, including chaos in many-body systems~\citep{maldacena2016remarks,nahum2017quantum,von2018operator,cotler2017chaos,fan2017out,khemani2018operator,page1993average,khemani2018operator,gu2017local}, and the  black-hole information paradox~\citep{hayden2007black,sekino2008fast,lashkari2013towards,gao2017traversable,shenker2014black,maldacena2017diving,roberts2015localized,shenker2014multiple,roberts2015diagnosing,blake2016universal,kitaev2014hidden}. 

One can analyze the scrambling effect by using the state representation of a quantum channel (also known as the Choi state), which encodes the input and output of a quantum channel into a quantum state~\citep{choi1975completely,jamiolkowski1972linear}. Within this formulation, quantum information scrambling can be measured by the tripartite mutual information (TMI) of a Choi state~\cite{hosur2016chaos,ding2016conditional,PhysRevE.98.052205,PhysRevA.97.042330,PhysRevLett.124.200504,PhysRevA.101.042324,PhysRevB.98.134303} which is written as
\begin{equation}
-I_3 = I(A:CD)-I(A:C)-I(A:D).\label{-I_3}
\end{equation}
Here, $A$ denotes a local region of the input subsystem whereas $C$ and $D$ are partitions of the output subsystem. The mutual information $I(A:X)$ quantifies the amount of information about $A$ stored in the region $X$. When $I(A:CD)>I(A:C)+I(A:D)$ or $-I_3>0$, it means that the amount of information about $A$ encoded in the whole output region $CD$ is larger than that in local regions $C$ and $D$. Therefore, $-I_3>0$ implies the delocalization of information or quantum information scrambling~\cite{hosur2016chaos}. Note that the TMI and the out-of-time-ordered correlator are closely related, suggesting that one can also use the out-of-time-ordered correlator as an alternative witness of quantum information scrambling~\citep{hosur2016chaos,landsman2019verified,swingle2016measuring,yao2016interferometric,zhu2016measurement,garttner2017measuring,swingle2018resilience,huang2019finite,yoshida2019disentangling,alonso2019out}.

From another point of view, because TMI is a multipartite entanglement measure, Eq.~\eqref{-I_3} can also be seen as a quantification of the multipartite \textit{entanglement in time}, i.e., the entanglement between input and output subsystems~\cite{hosur2016chaos}. Motivated by such an insight, one could expect that the scrambling effect can also be investigated from the perspective of temporal quantum correlations, i.e., temporal analogue of space-like quantum correlations. 

Moreover, Ku~\textit{et al.} \citep{ku2018hierarchy} has shown that three notable temporal quantum correlations (temporal nonlocality, temporal steering, and temporal inseperability) can be derived from a fundamental object called pseudo-density matrix~\citep{fitzsimons2015quantum,Ried2015,Zhao2018,Pi2019}, while elsewhere it was noted that there is a strong relationship between the Choi state and the pseudo-density matrix itself ~\citep{pisarczyk2019causal}. Taking inspiration from these connections, in this work, we aim to link the notion of scrambling to one particular scenario of temporal quantum correlation called temporal steering (TS)~\citep{chen2014temporal,chen2015detecting,chen2016quantifying,ku2016temporal,ku2018hierarchy,chen2017spatio,bartkiewicz2016temporal,liu2018quantum}.

Partly inspired by the Leggett-Garg inequality~\citep{leggett1985quantum,emary2013leggett}, temporal steering was developed as a temporal counterpart of the notion of spatial EPR steering~\citep{schrodinger1935discussion,wiseman2007steering,jones2007entanglement,cavalcanti2009experimental,piani2015necessary,skrzypczyk2014quantifying,costa2016quantification,branciard2012one,law2014quantum,uola2019quantum}. Recent work has shown that TS can quantify the information flow between different quantum systems~\citep{chen2016quantifying}, further suggesting it may also be useful in the study of scrambling.  Here, our goal is to demonstrate that \textit{one can witness information scrambling with temporal steering, which implies that the scrambling concept has nontrivial meaning in the broader context of temporal quantum correlations.} In addition, we wish to show that \textit{one can use ``measures'' developed to study temporal steering as a practical tool for the study of scrambling.}

We will restrict our attention to unitary channels of qubit systems, where the structure of non-scrambling channels can be well characterized~\cite{ding2016conditional}. More specifically, a unitary channel is non-scrambling, i.e. $-I_3=0$, if and only if the unitary is a ``criss-cross" channel that locally routes the local information from the input to the output subsystems. For qubit systems, a criss-cross channel can be described by a sequence of local unitaries and SWAP operations. 

The main result of this work is that we propose a quantity, $-T_3$, as a scrambling witness based on a measure of temporal steering called temporal steerable weight. We justify $-T_3$ to be a scrambling witness by proving that $-T_3=0$ when the global unitary channel is non-scrambling as mentioned above. We then compare the $-T_3$ with $-I_3$ by numerically simulating the Ising spin-chain model and the Sachdev-Ye-Kitaev (SYK) model. Finally, based on the one-sided device independent nature of steering, we point out that obtaining $-T_3$ requires less experimental resource than $-I_3$.

\begin{figure}
\includegraphics[width=0.95\linewidth]{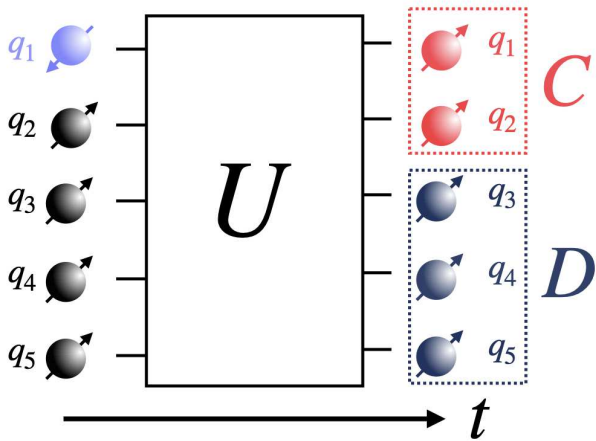}
\caption{\label{ets} Illustration of the extended temporal steering scenario involving 5 qubits labeled by $\{q_1,\cdots,q_5 \}$. Initially, Alice encodes her information in $q_1$ and lets the total system evolve. After the evolution, Bob divides the evolved system into two local regions $C$ and $D$, and tests the temporal steerability for each region to find out how the information spread throughout the whole system.}
\end{figure}

\section{Extended temporal steering scenario and scrambling witness}
\subsection{Temporal steering scenario}
Let us review the TS scenario~\cite{chen2014temporal} with the schematic illustration shown in Fig.~\ref{ets}. We focus on the reduced system $q_1$ and treat other qubits as the environment. In general, the TS task consists of many rounds of experiments. For each round, Alice receives $q_1$ with a fixed initial state $\rho_0$. Before the system evolves, Alice performs one of the projective measurements $\{E_{a|x}\}_{a,x}$ on $q_1$. Here, $x$ stands for the index of the measurement basis, to which Alice can freely choose,  and $a$ is the corresponding measurement outcome. The resulting post-measured conditional states can be written as
\begin{equation}
\Big\{\rho_{a|x}(0) = \frac{E_{a|x}\rho_0 E_{a|x}^{\dag} }{\tr(E_{a|x}\rho_0 E_{a|x}^{\dag})}\Big\}_{a,x},
\end{equation} where $\big\{p(a|x) = \tr(E_{a|x}\rho_0) \}_{a,x}$ predicts the probability of obtaining the outcome $a$ conditioned on Alice's choice $x$. Alice then sends the system to Bob through the quantum channel $\Lambda_t$, which describes the reduced dynamics of $q_1$ alone by tracing out other qubits. Finally, Bob performs another measurement on $q_1$ after the evolution. 

After all rounds of the experiments are finished, Alice sends her measurement results to Bob by classical communication, such that, Bob can also obtain the probability distribution $p(a|x)$. Additionally, based on the knowledge $(x,a)$ for each round of the experiment, Bob can approximate the conditional state $\rho_{a|x}(t)=\Lambda_t[\rho_{a|x}(0)]$ by quantum state tomography.  The aforementioned probability distribution and the conditional states can be summarized as a set called the TS assemblage $\{\sigma_{a|x}=p(a|x)\rho_{a|x}(t)\}_{a,x}$. Note that the assemblage can also be derived from the pseudo-density matrix (see Appendix~\ref{pdmchoi} for the derivation). In Appendix~\ref{pdmchoi}, we also show that the Choi matrix and the pseudo-density matrix are related by a partial transposition. 

Now, Bob can determine whether a given assemblage is steerable or unsteerable; that is, whether his system is quantum mechanically steered by Alice's measurements. In general, if the assemblage is unsteerable, it can be generated in a classical way, which is described by the local hidden state model:
\begin{equation}
\sigma_{a|x}^{\text{LHS}}(t)= \sum_{\lambda} p(a|x, \lambda) p(\lambda) \sigma_{\lambda}(t)~~~\forall a,x, \label{LHS}
\end{equation} where $\{p(\lambda), \sigma_\lambda(t)\}$ is an ensemble of local hidden states, and $\{p(a|x,\lambda)\}$ stands for classical post-processing. Therefore, the assemblage is steerable when it cannot be described by Eq.~\eqref{LHS}.

Bob can further quantify the magnitude of temporal steering~\cite{cavalcanti2016quantum,uola2019quantum}. Here, we use one of the quantifiers called temporal steerable weight (TSW)~\cite{chen2016quantifying}. For a given TS assemblage $\{\sigma_{a|x}(t)\}$, one can decompose it into a mixture of a steerable and unsteerable parts, namely,
\begin{equation}
\sigma_{a|x}(t) = \mu \sigma_{a|x}^\text{US}(t) +(1- \mu) \sigma_{a|x}^\text{S}(t) ~~\forall a,x, \label{decomposition}
\end{equation}
where $\{\sigma_{a|x}^\text{S}(t)\}$ and $\{\sigma_{a|x}^\text{US}(t)\}$ are the steerable and unsteerable assemblages, respectively, and $\mu$ stands for the portion (or weight) of the unsteerable part with $0\leq \mu\leq 1$. The TSW for the assemblage is then defined as 
\begin{equation}
\text{TSW}[\sigma_{a|x}(t)] = 1-\mu^*, \label{TSW}
\end{equation} 
where $\mu^*$ is the maximal unsteerable portion among all possible decompositions described by Eq.~\eqref{decomposition}. In other words, TSW can be interpreted as the minimum steerable resource required to reproduce the TS assemblage (e.g., $\text{TSW}=0$ for minimal steerability, and $\text{TSW}=1$ for maximal steerability). Note that Eq.~\eqref{TSW} can be numerically computed through semi-definite programming~\citep{cavalcanti2016quantum}.

According to Ref.~\cite{chen2016quantifying}, the TSW can reveal the direction of the information flow between an open quantum system and its environment during the time evolution.  When the information irreversibly flows out to the environment, TSW will monotonically decrease. Accordingly, the temporal increase of TSW implies information backflow. Recall that Alice steers $q_1$'s time evolution by her measurement $E_{a|x}$. In other words, the measurement encodes the information about $(a,x)$ in $q_1$. Therefore, after the evolution, Bob can estimate the amount of the information preserved in $q_1$ by computing the TSW.

\subsection{Extended temporal steering as a witness of scrambling}
As shown in Fig.~\ref{ets}, the evolution for the total system is still unitary, meaning that the information initially stored in $q_1$ is just redistributed (and localized) or scrambled after the evolution. Therefore, if we extend the notion of TS, which allows Bob to access the full system (regions $C$ and $D$), he can, in general, find out how the information localized or scrambled throughout the whole system. To be more specific, we now consider a global system with $N$ qubits labeled by $\{q_n\}_{n=1\dots N}$. Before Alice performs any measurement, we reset the total system by initializing the qubits in the maximally mixed state $\rho^{\text{tot}}_0 =  \mathbb{1}^{\otimes N}/2^N$, where $\mathbb{1}$ is the two-dimensional identity matrix. In this case, no matter how one probes the system, it gives totally random results, and no meaningful information can be learned. Then, Alice encodes the information $(a,x)$ in $q_1$ by performing $\{E_{a|x}\}$, which results in the conditional states of the total system:
\begin{align}
&\{\rho^{\text{tot}}_{a|x}(0) = \frac{1}{2^N}(2E_{a|x}\otimes \mathbb{1}^{\otimes N-1})\}_{a,x}\\
\text{with~~}&\{p(a|x) = \tr(E_{a|x}~\frac{\mathbb{1}}{2})=1/2 \}_{a,x}.
\end{align}
After that, let these conditional states evolve freely to time $t$, such that 
\begin{equation}
\rho^{\text{tot}}_{a|x}(t) = U_t\,\rho^{\text{tot}}_{a|x}(0)\,U_t^\dagger~~~\forall a,x,
\end{equation}
where $U_t$ can be any unitary operator acting on the total system. The assemblage for the global system then reads 
\begin{equation}
\{\sigma^{\text{tot}}_{a|x}(t) = p(a|x)\,\rho^{\text{tot}}_{a|x}(t)\}_{a,x}.
\end{equation}
Because the global evolution is unitary, it is straightforward that 
\begin{equation}
\text{TSW}[\sigma_{a|x}^\text{tot}(t)] = \text{TSW}[\sigma_{a|x}^\text{tot}(0)],
\end{equation}
which means that the information is never lost when all the degrees of freedom in the global system can be accessed by Bob.

In order to know how the information spread throughout all degrees of freedom, Bob can further analyze the assemblages obtained from different portions of the total system. For instance, he can divide the whole system into two local regions $C$ and $D$ as shown in Fig.~\ref{ets}, where $C$ contains $n_c$ qubits $\{q_1, \cdots, q_{n_c}\}$ and $D$ contains $n_d = N-n_c$ qubits $\{q_{n_c+1},\cdots ,q_N \}$, such that Bob obtains two additional assemblages: $\{\sigma_{a|x}^C (t) = \tr_D [\sigma_{a|x}^\text{tot}(t)]\}$ and 
$\{\sigma_{a|x}^D (t) = \tr_C [\sigma_{a|x}^\text{tot}(t)]\}$. Therefore, he can compute $\text{TSW}[\sigma_{a|x}^C (t)]$ and $\text{TSW}[\sigma_{a|x}^D (t)]$, estimating the amount of information localized in regions C and D. 

In analogy with Eq.~\eqref{-I_3}, we propose the following quantity to be a \textit{scrambling witness}: 
\begin{equation}
-T_3(t)  = \text{TSW}[\sigma_{a|x}^\text{tot}(t)] - \text{TSW}[\sigma_{a|x}^C(t)]-\text{TSW}[\sigma_{a|x}^D(t)],
\end{equation} where the minus sign for the quantity aims to keep the consistency with the TMI scrambling measure in Eq.~\eqref{-I_3}. It can be interpreted as the information stored in the whole system minus the information localized in regions $C$ and $D$; namely, the information scrambled to the non-local degrees of freedom. 

As mentioned in the introduction section, for a non-scrambling channel consisting of local unitaries and SWAP operations, the information will stay localized (non-scrambled). Therefore, we further justify that $-T_3(t)$ can be a scrambling witness, under the assumption of global {\em unitary} evolution, by proving that under non-scrambling evolutions, this quantity will vanish, i.e. $-T_3=0$. Accordingly, any nonzero value of $-T_3$ can be seen as a witness of scrambling.
\begin{theorem}
If the global unitary evolution $U$ is local for regions $C$ and $D$, that is, $U = U_C\otimes U_D$, the resulting $-T_3$ is zero.\label{local}
\end{theorem}

The proof is given in Appendix~\ref{pf:1}.

\begin{theorem}
If the global unitary $U$ is a SWAP operation between qubits, then $-T_3(t) = 0$.\label{swap}
\end{theorem}

The proof is given in Appendix~\ref{pf:2}.

According to the results of Theorem~\ref{local} and Theorem~\ref{swap}, we conclude that $-T_3(t)$ will vanish if the global evolution is any sequence of local unitaries and SWAP operations, as required for a witness of scrambling.

\section{numerical simulations}
In this section, we present the numerical simulations for the Ising spin chain and the SYK model. For simplicity, we consider $\{E_{a|x}\}$ to be projectors of Pauli matrices $\{\sigma_x,~\sigma_y,~\sigma_z\}$ such that $\text{TSW}[\sigma_{a|x}^\text{tot}(0)]=1$~\cite{chen2016quantifying}.

\subsection{Example 1: The Ising spin chain}

\begin{figure*}
\includegraphics[width=0.8\paperwidth]{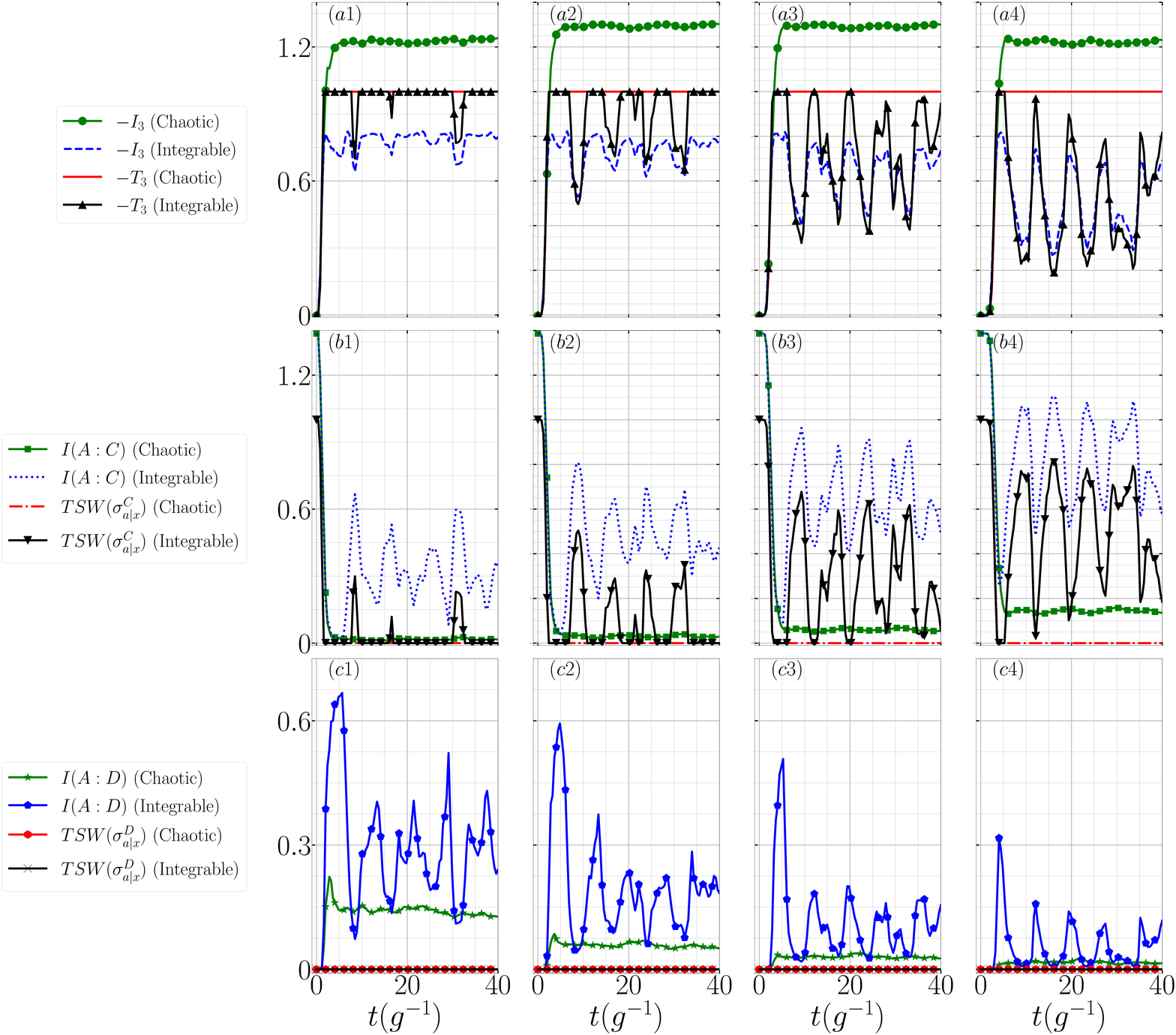}
\caption{\label{spin_scrambling} Information scrambling and localization with different output partitions [ ($n_c=2$, $n_d=5$) for (a1, b1, c1); ($n_c=3$, $n_d=4$) for (a2, b2, c3); ($n_c=4$, $n_d=3$) for (a3, b3, c3); ($n_c=3$, $n_d=4$) for (a4, b4, c4)] for the chaotic ($g=1$, $h=0.5$) and the integrable ($g=1$, $h=0$) spin chain dynamics. (a) Information scrambling measured by $-I_3$ and witnessed by $-T_3$. (b) Information stored in region $C$ and measured by $I(A:C)$ and TSW($\{\sigma_{a|x}^C\}$). (c) Information stored in region $D$ and measured by $I(A:D)$ and TSW($\{\sigma_{a|x}^D\}$). Here, $n_c$ and $n_d$ denote the number of qubits involved in region $C$ and region $D$, respectively.}
\end{figure*}

\begin{figure*}
\includegraphics[width=0.8\paperwidth]{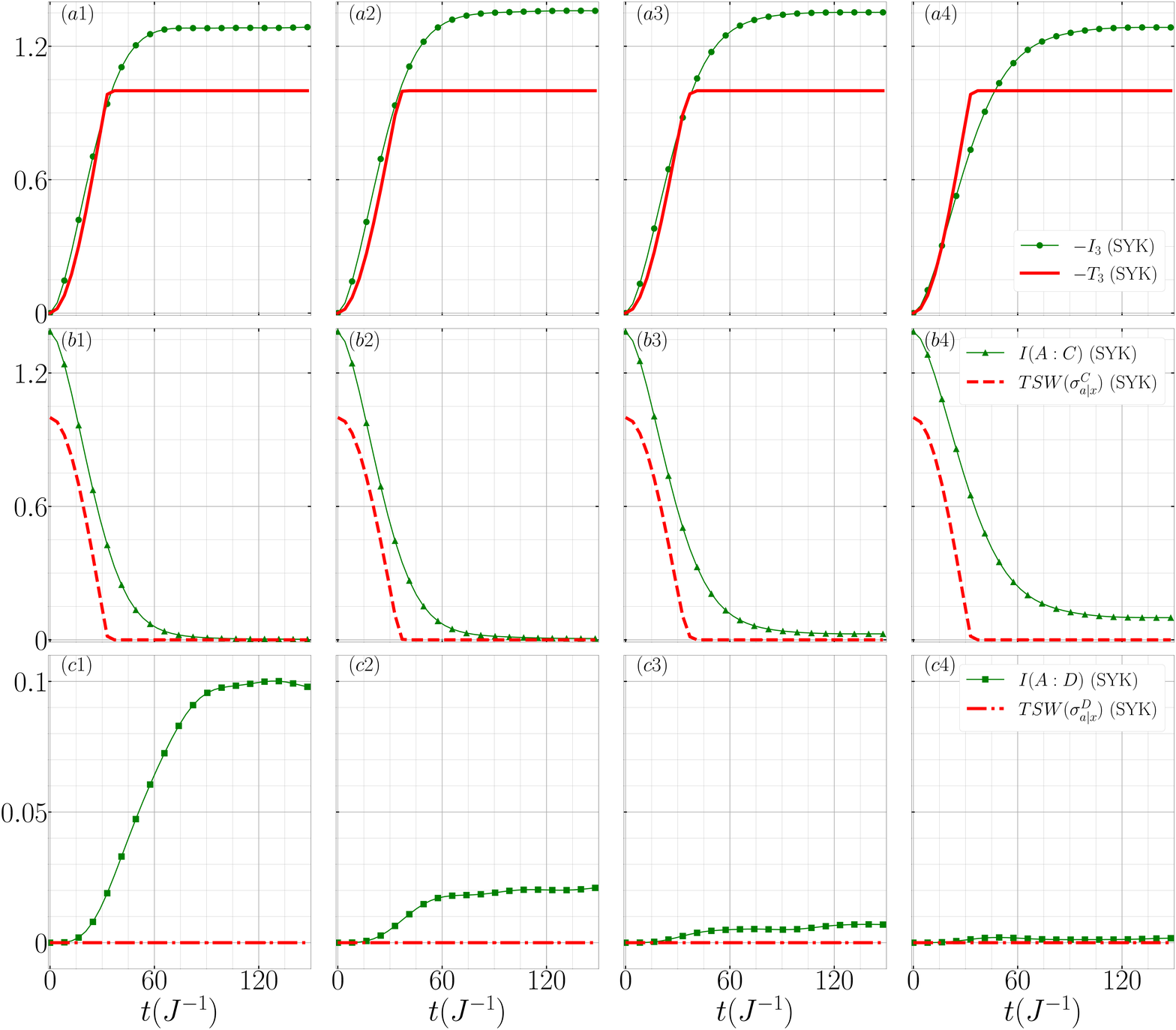}
\caption{\label{syk_scrambling} Information scrambling and localization of the Sachdev-Ye-Kitaev model with different partitions of the output system [ ($n_c=2$, $n_d=5$) for (a1, b1, c1); ($n_c=3$, $n_d=4$) for (a2, b2, c3); ($n_c=4$, $n_d=3$) for (a3, b3, c3); ($n_c=3$, $n_d=4$) for (a4, b4, c4)]. (a) Information scrambling measured by $-I_3$ and witnessed $-T_3$. (b) Information localized in region $C$ and measured by $I(A:C)$ and TSW($\sigma_{a|x}^C$). (c) Information localized in region $D$ and measured by $I(A:D)$ and TSW($\sigma_{a|x}^D$).}
\end{figure*}

We now consider a one-dimensional Ising model of $N$ qubits with the Hamiltonian
\begin{equation}
H = - \sum_{i=1}^{N-1}\sigma_{i}^{z}\sigma_{i+1}^{z} - h\sum_{i=1}^{N}\sigma_i^z - g\sum_{i=1}^{N}\sigma_i^x.
\end{equation}
The key feature is that one can obtain chaotic behavior by simply turning on the longitudinal field parametrized by $h$. 

Here, we consider the system containing 7 qubits $\{q_1\dots q_7\}$ and compare the dynamical behavior of information scrambling for chaotic ($g=1$, $h=0.5$) and integrable regimes ($g=1$, $h=0$) by encoding the information in $q_1$. 

As shown in Fig.~\ref{spin_scrambling}, we plot the information scrambling measured by $-I_3$ and witnessed by $-T_3$ and the amount of information stored in region $C$ ($D$) with the quantities $I(A:C)$ and TSW$(\sigma_{a|x}^C)$ [$I(A:D)$ and TSW$(\sigma_{a|x}^D)$] for different partitions of the output system. For a fixed output partition, [Fig. ~\ref{spin_scrambling}(a3), ~\ref{spin_scrambling}(b3), ~\ref{spin_scrambling}(c3) for instance], one can find that the local minima of the scrambling corresponds to the local maxima of the information stored either in region $C$ or region $D$. Therefore, we can conclude that \textit{the decrease of the scrambling during the evolution results from the information backflow from non-local degrees of freedom to local degrees of freedom.}

Moreover, information scrambling behaves differently for chaotic and integrable evolutions. For chaotic evolution, the scrambling will remain large after a period of time, because the information is mainly encoded in non-local degrees of freedom. However, for integrable systems, we can observe that both $-I_3$ and $-T_3$ show oscillating behavior. Furthermore, as the dimension of region $C$ becomes larger, the oscillating behavior of the scrambling for integrable cases significantly increases, whereas the scrambling patterns for chaotic cases remain unchanged.

\subsection{Example 2 : The Sachdev-Ye-Kitaev model}
We now consider the SYK model which can be realized by a Majorana fermionic system with the Hamiltonian
\begin{equation}
\begin{aligned}
H = \sum_{i<j<k<l} J_{ijkl} \chi_i \chi_j \chi_k \chi_l \: , \\
\overline{J_{ijkl}^2} = \frac{3!}{(N-1)(N-2)(N-3)} J^2 ,
\end{aligned}
\end{equation}
where the $\chi_i$ represent Majorana fermions with $j,k,l,m=1,...N$.
Meanwhile, $J_{ijkl}$ in the Hamiltonian follow the random normal distribution with zero mean and variance $\overline{J_{ijkl}^2}$. To study this model in qubit system, we can use the Jordan-Wigner transformation
\begin{align}
\chi_{2i-1} &= \frac{1}{\sqrt{2}} X_1 X_2 ... X_{i-1} Z_i , \nonumber\\ \chi_{2i} &= \frac{1}{\sqrt{2}} X_1 X_2 ... X_{i-1} Y_i
\end{align}

to convert the Majorana fermions to spin chain Pauli operators. In our numerical results, we consider $N=14$ (a seven qubits system) and $J=1$.
Figure~\ref{syk_scrambling} shows the time evolutions of the information scrambling and the information localized in region $C$ and $D$ for different partitions of the output system similar to those in Example $1$.

The main difference between these examples is that, in Example $1$, the qubits only interact with their nearest neighbors; whereas in example $2$, the model includes the interactions to all other qubits. Therefore, we can observe that in the spin chain model, the scrambling is sensitive to the dimension of the output system. However, in the SYK model, the scrambling is not susceptible to the partition of the output system, namely, the scrambling time and the magnitude of the tripartite mutual information after the scrambling period of different output partitions are similar (asymptotically reaching the Harr-scrambled value~\cite{hosur2016chaos}). Note that in Appendix~\ref{finitesize}, we also provide numerical simulations involving different number of qubits for the above examples. We find that when decreasing (increasing) the number of qubits, the tendency of information backflow~\cite{chen2016quantifying} from global to local degrees of freedom will increase (decrease) for both chaotic and integrable dynamics.

Finally, for the scrambling dynamics (chaotic spin chain and SYK model), we can find that $\text{TSW}(\sigma_{a|x}^C)$ degrades more quickly to zero than $I(A:C)$. In addition, $\text{TSW}(\sigma_{a|x}^D)$ remains zero all the time, while $I(A:D)$ could reach some non-zero value. The different behavior between the $\text{TSW}(\sigma_{a|x}^{C/D})$ and $I(A:C/D)$ results from the hierarchical relation between these two quantities~\citep{ku2018hierarchy}, which states that temporal quantum steering is a stricter quantum correlation than bipartite mutual information. In other words, we can find some moments where $I(A:C/D)$ has non-zero value whereas $\text{TSW}(\sigma_{a|x}^{C/D})$ is zero, but not \textit{vice versa}. 
The situation when $\text{TSW}(\sigma_{a|x}^{C})=\text{TSW}(\sigma_{a|x}^{D})=0$ [$I(A:C)=I(A:D)=0$] implies that $-T_3$ [$-I_3$] reaches its maximum. Therefore, for scrambling dynamics we can observe that $-T_3$ reaches its maximum earlier than $-I_3$.

\section{Summary}
\begin{table*}
\begin{ruledtabular}
\begin{tabular}{lll}
  & Space-like structure & Time-like structure\\
\hline Diagram&&\\[15pt]
 & \raisebox{0pt}{\includegraphics[width=0.2\paperwidth]{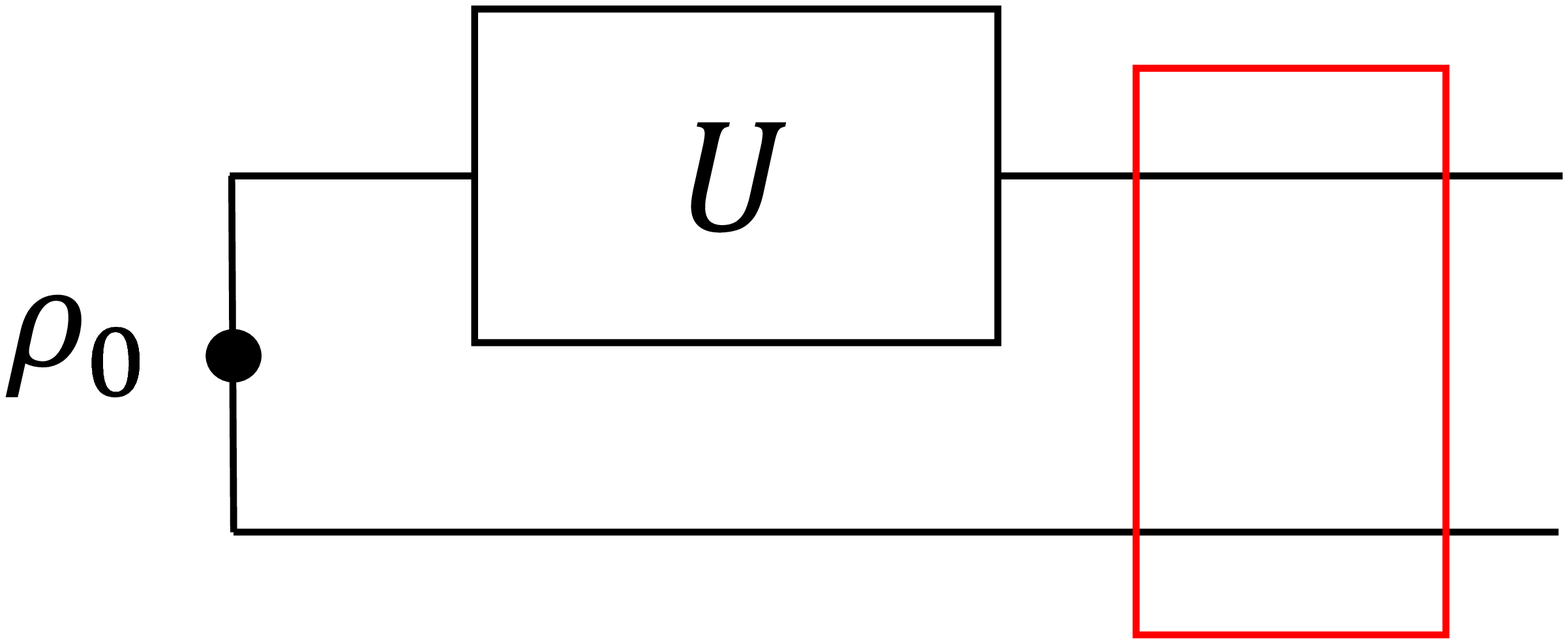}} &\raisebox{0pt}{\includegraphics[width=0.2\paperwidth]{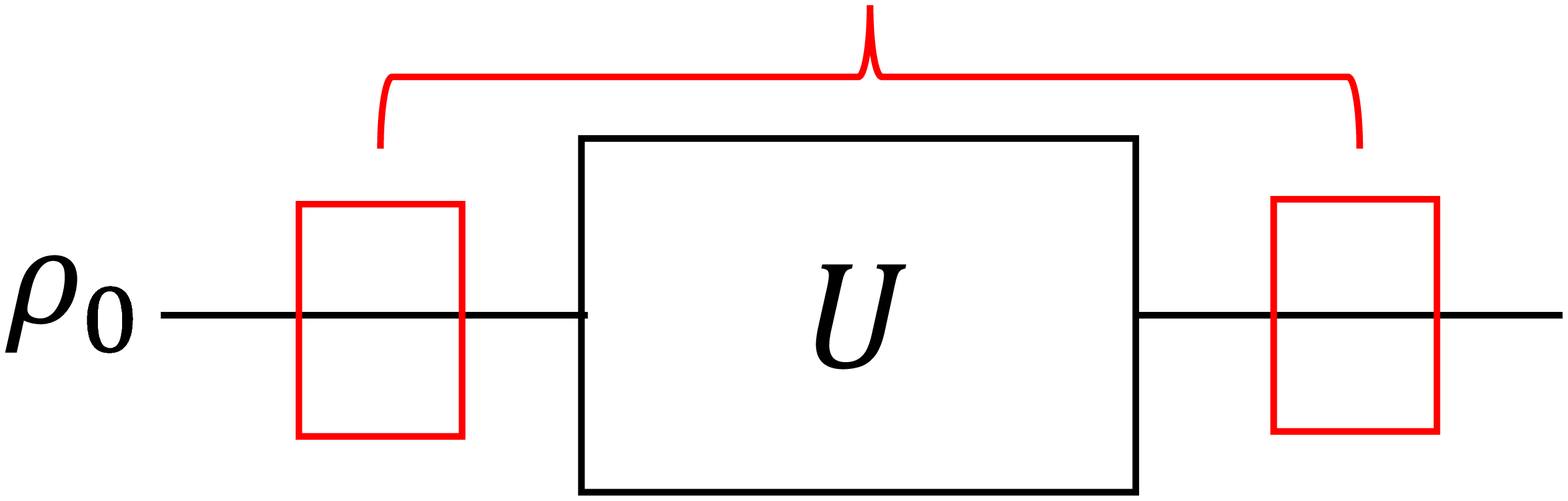}}\\
\hline
Operator & Choi Matrix & Pseudo Density Matrix \\
\hline
\vtop{\hbox{\strut Quantum}\hbox{\strut correlations}} & \vtop{\hbox{\strut Mutual information}\hbox{\strut Entanglement}\hbox{\strut CHSH inequality}} & \vtop{\hbox{\strut Temporal steering}\hbox{\strut Temporal inseparability}\hbox{\strut Legget-Garg inequality}} \\[30pt]
\hline
\vtop{\hbox{\strut Input} \hbox{\strut state $\rho_0$}} & EPR pairs & Maximally mixed states \\
\end{tabular}
\end{ruledtabular}
\caption{\label{sttable} Relations between space-like and time-like structures. In the diagrams, the vertical line with a dot in the middle for the space-like structure represents the bipartite entanglement as the resource of the input state. The red boxes represent the measurements required for the spatial/temporal quantum state tomography for different scenarios.}
\end{table*}

In summary, we demonstrate that the information scrambling can not only be verified by the spatial quantum correlations encoded in a Choi matrix but also the temporal quantum correlations encoded in a pseudo-density matrix (see Table ~\ref{sttable} for the comparison between the space-like and time-like  structures). Moreover, we further provide an information scrambling witness, $-T_3$, based on the extended temporal steering scenario.

A potential advantage of using $-T_3$ as a scrambling witness, over $-I_3$, is that $-T_3$ requires less measurement resources than $-I_3$. More specifically, when measuring $-T_3$, we do not have to access the full quantum state of the input region $A$, because in the steering scenario Alice's measurement bases are only characterized by the classical variable $x$. From a practical point of view, the number of Alice's measurement basis can be less than that required for performing quantum state tomography on the region $A$. For the examples presented in this work, we consider that region $A$ only contains a single qubit ($q_1$), in which the standard choice of the measurement bases is the set of Pauli matrices, $\{\sigma_x, \sigma_y,\sigma_z\}$. For the steering scenario, we can choose only two of these matrices as Alice's measurement bases, though for the numerical simulations presented in this work, we still consider that all three Pauli matrices are used by Alice. 

Once the dimension of region $A$ increases, the number of the measurements required to perform quantum state tomography and obtain $-I_3$ will also increase. However, as aforementioned, for the steering scenario, the dimension of the region $A$ is not assumed, implying that we can still choose a manageable number of Alice's measurements to verify the steerability and compute $-T_3$.

Finally, it is important to note that we only claim that $-T_3$ is a witness of scrambling rather than a quantifier, because we only prove that $-T_3$ vanishes whenever the evolution is non-scrambling. An open question immediately arises: Can $-T_3$ be further treated as a quantifier from the viewpoint of resource theory~\citep{chitambar2019quantum}? To show this, our first step would be to prove that $-T_3$ monotonically decreases whenever the evolution is non-scrambling, and we leave it as a future work.

\textit{Note added}---After this work was completed, we became aware of \citep{zhang2020quantum}, which independently showed that the temporal correlations are connected with information scrambling, because the out-of-time-ordered correlators can be calculated from pseudo-density matrices.
\begin{acknowledgments}
This work is supported partially by the National Center for Theoretical Sciences and Ministry of Science and Technology, Taiwan, Grants No. MOST 107-2628-M-006-002-MY3, and MOST 109-2627-M-006 -004, and the Army Research Office (under Grant No. W911NF-19-1-0081). N.L. acknowledges partial support from JST PRESTO through Grant No. JPMJPR18GC, the Foundational Questions Institute (FQXi), and the NTT PHI Laboratory. F.N. is supported in part by: Nippon Telegraph and Telephone Corporation (NTT) Research, the Japan Science and Technology Agency (JST) [via the Quantum Leap Flagship Program (Q-LEAP), the Moonshot R\&D Grant Number JPMJMS2061, and the Centers of Research Excellence in Science and Technology (CREST) Grant No. JPMJCR1676], the Japan Society for the Promotion of Science (JSPS) [via the Grants-in-Aid for Scientific Research (KAKENHI) Grant No. JP20H00134 and the JSPS–RFBR Grant No. JPJSBP120194828], the Army Research Office (ARO) (Grant No. W911NF-18-1-0358), the Asian Office of Aerospace Research and Development (AOARD) (via Grant No. FA2386-20-1-4069), and the Foundational Questions Institute Fund (FQXi) via Grant No. FQXi-IAF19-06.
\end{acknowledgments}

\appendix
\section{Relation between Choi matrix and pseudo density matrix \label{pdmchoi}}
To illustrate the main idea behind the TMI scrambling measure in Ref.~\citep{hosur2016chaos}, let us now consider a system made up of $N$ qubits, labeled by $\{q_1,\cdots,q_N\}$, with a Hilbert space $\mathcal{H}^\text{In}_q = \bigotimes_{i=1}^N\mathcal{H}^{\text{In}}_{q_i}$. We then create $N$ ancilla qubits, labeled with $\{\tilde{q}_1,\cdots,\tilde{q}_N\}$, where  each $\tilde{q}_i$ is maximally entangled with the corresponding qubit $q_i$. Therefore, the Hilbert space of the total $2N$ qubits system is $\mathcal{H}_{\tilde{q}}^\text{In}\otimes \mathcal{H}_q^\text{In}$. The corresponding density operator is $\rho_0^\text{CJ} \in L(\mathcal{H}_{\tilde{q}}^\text{In})\otimes L(\mathcal{H}_q^\text{In})$, where $L(\mathcal{H}_q^\text{X})$ denotes the set of linear operators on the Hilbert state $\mathcal{H}_q^\text{X}$. We can expand $\rho_0^\text{CJ}$ with Pauli matrices such that
\begin{equation}
\rho^{CJ}_0 = \frac{1}{4^N} \sum_{i_1,\cdots i_N=0}^3 T_{i_1\cdots i_N}(\bigotimes_{m=1}^N \sigma_{i_m})\otimes (\bigotimes_{m=1}^N \sigma_{i_m}) ,
\end{equation}
where $T_{\mu_1\cdots\mu_N} = V_{\mu_1}\cdots V_{\mu_N}$, $\mathbf{V} = (+1,+1,-1,+1)$, and $\bm{\sigma} = (\mathbb{1},\sigma_x,\sigma_y,\sigma_z) $.
Let us now send the original qubits into a quantum channel (completely positive and trance preserving map) $\Phi_t: L(\mathcal{H}^{\text{In}}_q)\rightarrow L(\mathcal{H}^{\text{Out}}_q)$. Here, we consider the channel to be unitary; namely, $\Phi_t(\rho) = U_t \rho U_t^\dagger$, where $U_t$ is a unitary operator. The evolved density matrix (known as the Choi matrix) then reads
\begin{align}
\rho^{CJ}_t &= (\mathbb{1}\otimes\Phi_t)[\rho_0^\text{CJ}]\nonumber\\&=(\mathbb{1}\otimes U_t )\rho^{CJ}_0 (\mathbb{1}\otimes U_t^\dagger) \in L(\mathcal{H}_{\tilde{q}}^\text{In}) \otimes L(\mathcal{H}_q^\text{Out}).
\end{align}In general, $U_t$ can be expanded as
\begin{equation}
U_t=\sum_{\mu_1\cdots\mu_N}u_{\mu_1\cdots\mu_N}\bigotimes_{m=1}^N \sigma_{\mu_m}.
\end{equation}
We therefore can expand the Choi matrix into:
\begin{flalign}
\rho^{CJ}_t&=\frac{1}{4^N}\sum_{i_1\cdots i_N}\sum_{j_1\cdots j_N} \Omega_{j_1\cdots j_N}^{i_1\cdots i_N}(\bigotimes_{m}^N \sigma_{i_m}) \otimes (\bigotimes_{n}^N \sigma_{j_n}),& \nonumber\\
\Omega_{j_1\cdots j_N}^{i_1\cdots i_N}&=
\frac{1}{2^N}\sum_{\mu_1\cdots\mu_N}\sum_{\nu_1\cdots\nu_N}[T_{i_1\cdots i_N}u_{\mu_1\cdots \mu_N}u^*_{\nu_1\cdots \nu_N}\times&\nonumber\\
&\qquad\qquad\qquad\qquad~\prod_{m=1}^N \tr (\sigma_{j_m}\sigma_{\mu_m}\sigma_{i_m}\sigma_{\nu_m})]&\label{CJ}.
\end{flalign}

We now construct the pseudo-density matrix (PDM) through a \textit{temporal analogue of quantum state tomography (QST)} between measurement events at two different moments~\citep{fitzsimons2015quantum}. A PDM for an $N$ qubits system in an initially maximally mixed state undergoing $\Phi_t$ is given by
\begin{align}
R_t&=\frac{1}{4^N}\sum_{i_1\cdots i_N}\sum_{j_1\cdots j_N} C_{j_1\cdots j_N}^{i_1\cdots i_N}(\bigotimes_{m}^N \sigma_{i_m}) \otimes (\bigotimes_{n}^N \sigma_{j_n}),\nonumber\\
C_{j_1\cdots j_N}^{i_1\cdots i_N}&=\mathbf{E}[~\{\bigotimes_{m}^N \sigma_{i_m} , \bigotimes_{n}^N \sigma_{j_n}\}~]\nonumber\\
&= \frac{1}{2^N}\sum_{\mu_1\cdots\mu_N}\sum_{\nu_1\cdots\nu_N} [u_{\mu_1\cdots \mu_N}u^*_{\nu_1\cdots \nu_N}\times\nonumber\\
&~~~~~~~~~~~~~~~~~~~~~~~~~~\prod_{m=1}^N \tr (\sigma_{j_m}\sigma_{\mu_m}\sigma_{i_m}\sigma_{\nu_m})],\label{PDM}
\end{align}
where $\mathbf{E}[~\{\bigotimes_{m}^N \sigma_{i_m} , \bigotimes_{n}^N \sigma_{j_n}\}~]$ is the expectation value of the product of the outcome of the measurement $\bigotimes_{m}^N \sigma_{i_m}$ performed on the initial time and the outcome of the measurement $\bigotimes_{n}^N \sigma_{j_n}$ performed at the final time $t$. Similarly, $R_t \in L(\mathcal{H}_{\tilde{q}}^\text{In}) \otimes L(\mathcal{H}_q^\text{Out}) $.

By comparing the coefficients of the $N$ qubits Choi matrix ($\Omega_{j_1\cdots j_N}^{i_1\cdots i_N}$) in Eq.~\eqref{CJ} with those of the PDM in Eq.~\eqref{PDM} ($C_{j_1\cdots j_N}^{i_1\cdots i_N}$), one can find that these two matrices are related through a partial transposition of the input degree of freedom, i.e.
\begin{equation}
(\rho^\text{CJ}_t)^{T_\text{In}} = R_t.
\label{CJPDM}
\end{equation} 

According to Ref.~\citep{ku2018hierarchy}, the TS assemblage can also be derived from the pseudo density matrix $R_t$ [which is defined in Eq.~\eqref{PDM}] by the following Born's rule:
 \begin{equation}
 \sigma_{a|x}(t) = \tr_{\text{In}}[(E_{a|x}\otimes \mathbb{1}^{\otimes 2N-1})R_t],
 \end{equation}
 where $\tr_{\text{In}}$ denotes the partial trace over the input Hilbert space.

As mentioned in the main text, the notion of
scrambling can be understood as the multipartite entanglement in the Choi state. Therefore, the insight
inferred from Eq.~\eqref{CJPDM} suggests us that it should be possible to reformulate the information scrambling with multipartite temporal quantum correlations.

\section{Proof of Theorem 1\label{pf:1}}
\begin{proof}
Let's start from the evolved assemblage for the total system (region $CD$):
\begin{align}
\sigma^\text{tot}_{a|x}(t) &= U_C\otimes U_D \Big[\frac{1}{2^N}(E_{a|x}\otimes\mathbb{1}^{\otimes N-1})\Big] U_C^\dagger \otimes U_D^\dagger \nonumber\\
&=U_C\Big[\frac{1}{2^{n_c}}(E_{a|x}\otimes\mathbb{1}^{\otimes n_c-1})\Big]U_C^\dagger \otimes U_D \frac{\mathbb{1}^{\otimes n_d}}{2^{n_d}}U_D^\dagger,\\
\sigma_{a|x}^C(t) &= U_C\Big[\frac{1}{2^{n_c}}(E_{a|x}\otimes\mathbb{1}^{\otimes n_c-1})\Big]U_C^\dagger,\\
\sigma_{a|x}^D(t) &= U_D \frac{\mathbb{1}^{\otimes n_d}}{2^{n_d+1}}U_D^\dagger.
\end{align}
Since $U_C$ and $U_D$ are unitary, leading to the invariance of the TSW, we find the following results:
\begin{align}
\text{TSW}[\sigma_{a|x}^\text{tot}(t)] &= \text{TSW}[\sigma_{a|x}^\text{tot}(0)] = \text{TSW}\Big(\frac{E_{a|x}\otimes \mathbb{1}^{\otimes N-1}}{2^N}\Big),\\
\text{TSW}[\sigma_{a|x}^C(t)] &= \text{TSW}[\sigma_{a|x}^C(0)] = \text{TSW}\Big( \frac{E_{a|x}\otimes \mathbb{1}^{\otimes n_c-1}}{2^{n_c}}\Big),\\
\text{TSW}[\sigma_{a|x}^D(t)] &= \text{TSW}[\sigma_{a|x}^D(0)] = \text{TSW}\Big( \frac{\mathbb{1}^{\otimes n_d}}{2^{n_d+1}}\Big)
\end{align}
It is straightforward to conclude that $\text{TSW}[\sigma_{a|x}^D(0)] = 0$, since $\{\sigma_{a|x}^D(0)\}$ can be decomposed as the local hidden state model shown in Eq.~\eqref{LHS}. In addition,
\begin{equation}
\text{TSW}\Big(\frac{E_{a|x}\otimes \mathbb{1}^{\otimes n-1}}{2^{n}}\Big) = \text{TSW}\Big(\frac{E_{a|x}}{2}\Big)
\end{equation} for arbitrary positive integer $n$. Therefore, we can deduce that 
\begin{equation}
-T_3(t) = \text{TSW}[\sigma_{a|x}^\text{tot}(t)]-\text{TSW}[\sigma_{a|x}^C(t)]-\text{TSW}[\sigma_{a|x}^D(t)]=0.
\end{equation}

\end{proof}

\section{Proof of Theorem 2 \label{pf:2}}
\begin{proof}
We can find that the sum of the TSW for regions $C$ and $D$ is invariant under any permutation between qubits such that
\begin{align}
&\text{TSW}[\sigma_{a|x}^C(t)]+\text{TSW}[\sigma_{a|x}^D(t)]\nonumber \\
 = &\text{TSW}\Big(\frac{E_{a|x}}{2}\Big)+\text{TSW}\Big(\frac{\mathbb{1}}{4}\Big)
\end{align}
Therefore, under the SWAP operation, $-T_3(t) = \text{TSW}(\frac{E_{a|x}}{2})- \text{TSW}(\frac{E_{a|x}}{2})=0$.
\end{proof}

\section{The qubit Clifford scrambler}
\begin{figure*}
\includegraphics[width=1\linewidth]{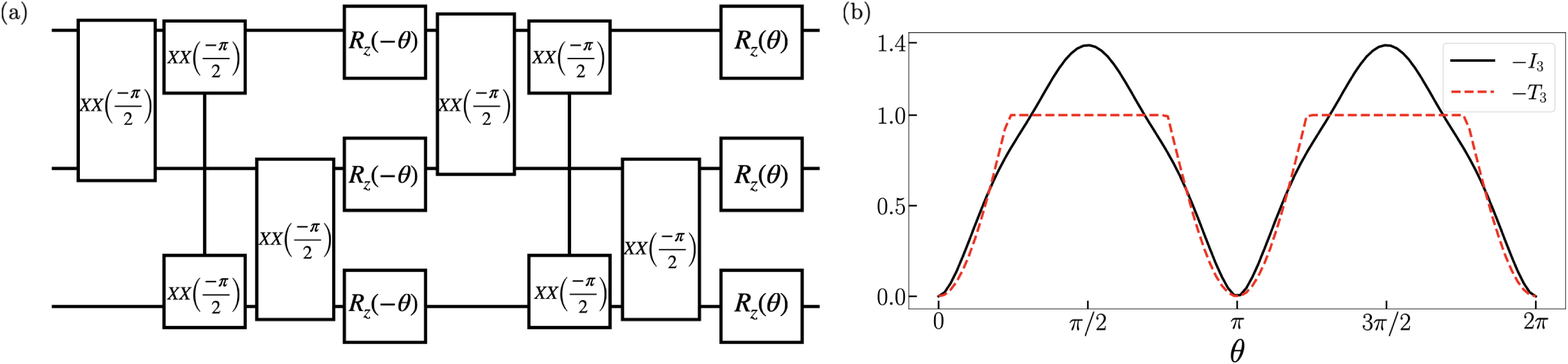}
\caption{\label{Clifford} (a) The circuit diagram of the Clifford scrambling circuit, where $XX$ stands for the Ising ($XX$) coupling and $R_z$ stands for the rotation-$z$ gate. One can obtain different degrees of scrambling by changing the angle $\theta$: $\theta=0$ for the non-scrambling case and $\theta=\pi/2\pm n \pi$ for the maximum scrambling case. Here, $n$ is an arbitrary integer. (b) Numerical simulations of $-I_3$ (black solid) and $-T_3$ (red dashed) for the Clifford scrambler for different angles $\theta$.}
\end{figure*}

\begin{figure*}
\includegraphics[width=1.1\linewidth]{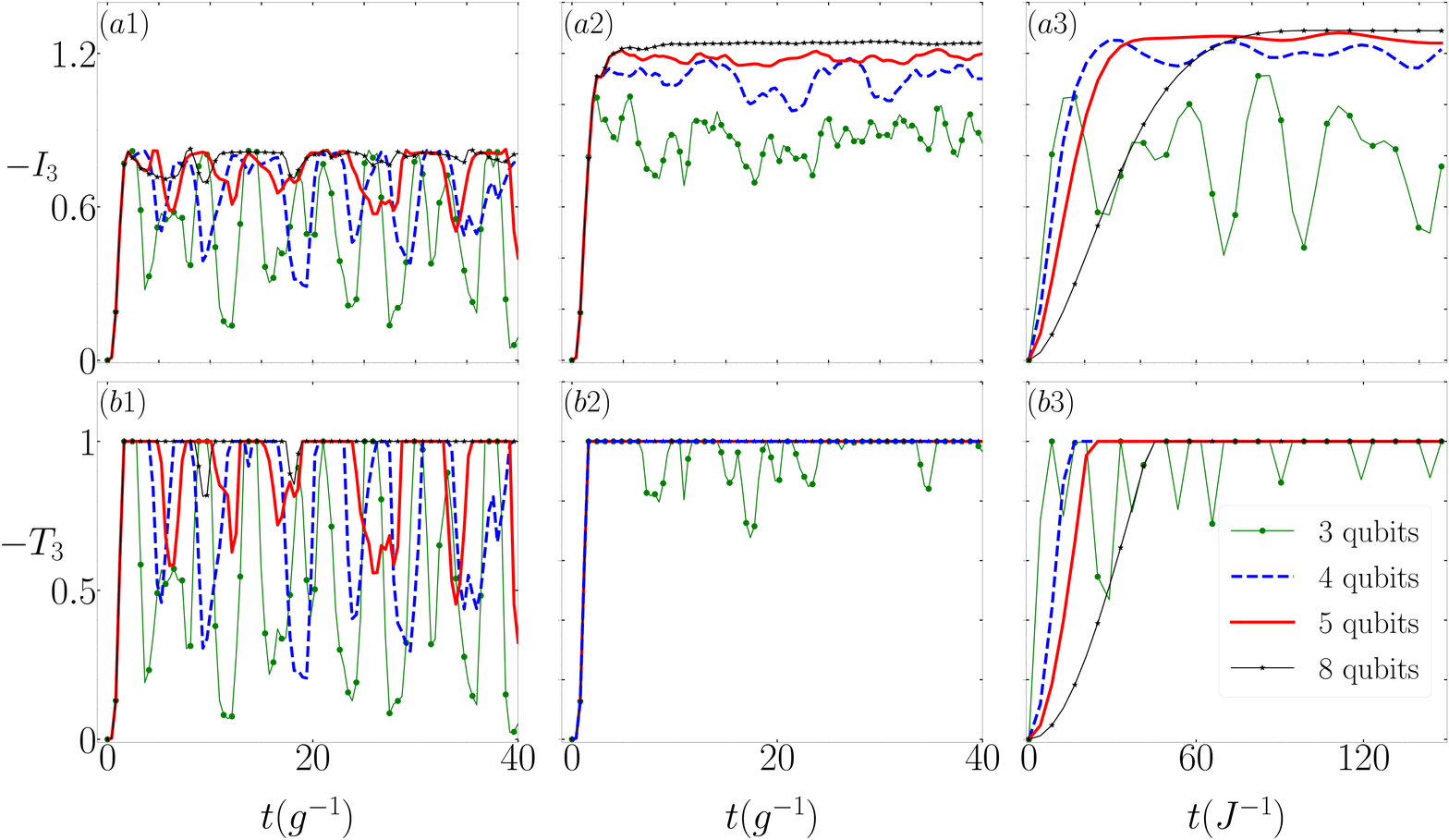}\\
\caption{\label{finite} In this figure we numerically simulate $-I_3$ and $-T_3$ of the integrable spin chain (a1, b1), the chaotic spin chain (a2, b2), and the SYK model (a3, b3), for different number of qubits. We find that (for both the integrable and chaotic systems) the oscillations of information scrambling are enhanced when the system size decreases. It suggests that the tendency of information backflow from non-local to local degrees of freedom increases when the system size decreases. Note that the numbers of qubits in region $C$ for 3-qubit, 4-qubit, 5-qubit and 8-qubit systems are 1, 2, 3, and 4, respectively.}
\end{figure*}
In this section, we numerically analyze the qubit Clifford scrambling circuit, proposed in Ref.~\cite{landsman2019verified}. The setting only involves three qubits with a quantum circuit depicted in Fig.~\ref{Clifford}, which is parametrized by $\theta$. By changing the angle $\theta$, one can scan the angle from non-scrambling ($\theta=0$) to maximally scrambling ($\theta = \pm \frac{\pi}{2}$), which can be described by the following unitary matrix
\begin{equation}
U_s=\frac{i}{2}
\begin{pmatrix}
-1&0&0&-1&0&-1&-1&0\\
0&1&-1&0&-1&0&0&1\\
0&-1&1&0&-1&0&0&1\\
1&0&0&1&0&-1&-1&0\\
0&-1&-1&0&1&0&0&1\\
1&0&0&-1&0&1&-1&0\\
1&0&0&-1&0&-1&1&0\\
0&-1&-1&0&-1&0&0&-1
\end{pmatrix}.
\end{equation}
According to Ref.~\cite{landsman2019verified}, the scrambling unitary delocalizes all single qubit Pauli operators to three qubit Pauli operators in the following way:
\begin{align}
U_s(\sigma_x\otimes \mathbb{1}\otimes\mathbb{1})U_s^\dagger &=- \sigma_x\otimes \sigma_y\otimes \sigma_y\nonumber\\
U_s(\sigma_y\otimes \mathbb{1}\otimes\mathbb{1})U_s^\dagger &=- \sigma_y \otimes \sigma_z \otimes \sigma_z \nonumber\\
U_s(\sigma_z\otimes \mathbb{1}\otimes\mathbb{1})U_s^\dagger &=- \sigma_z \otimes \sigma_x \otimes \sigma_x\nonumber\\
U_s(\mathbb{1}\otimes \sigma_x\otimes\mathbb{1})U_s^\dagger &=- \sigma_y \otimes \sigma_x \otimes \sigma_y \nonumber\\
U_s(\mathbb{1}\otimes \sigma_y\otimes\mathbb{1})U_s^\dagger &=- \sigma_z \otimes \sigma_y \otimes \sigma_z\nonumber\\
U_s(\mathbb{1}\otimes \sigma_z\otimes\mathbb{1})U_s^\dagger &=- \sigma_x \otimes \sigma_z \otimes \sigma_x\nonumber\\
U_s(\mathbb{1}\otimes \mathbb{1}\otimes\sigma_x)U_s^\dagger &=- \sigma_y\otimes \sigma_y \otimes\sigma_x\nonumber\\
U_s(\mathbb{1}\otimes \mathbb{1}\otimes\sigma_y)U_s^\dagger &=- \sigma_z\otimes \sigma_z \otimes \sigma_y\nonumber\\
U_s(\mathbb{1}\otimes \mathbb{1}\otimes\sigma_z)U_s^\dagger &=- \sigma_x  \otimes \sigma_x \otimes \sigma_z.
\end{align}Such a delocalization is often known as operator growth, which can be viewed as a key signature of quantum scrambling. 
In Fig.~\ref{Clifford}, we plot the values of $-T_3$ and $-I_3$ by changing the angles $\theta$. We can see that both $-I_3$ and $-T_3$ display an oscillating pattern with period $\pi$. The value of $-I_3$ reaches its maximum scrambling value at $\theta=\pi/2$; while, $-T_3$ reaches its maximum scrambling value earlier than $-I_3$ due to the sudden vanishing of the TSW for local regions.

\section{Numerical simulations for different system sizes  \label{finitesize}}
\renewcommand{\arraystretch}{1.5}
\begin{table}
\begin{tabular}{ |c|c|c|c|c| }
\hline
$\mathcal{I}_{I_3}(T)$ & 3-qubit & 4-qubit & 5-qubit & 8-qubit\\[5pt]
\hline
Spin chain(Integrable) & 5.295 & 2.602 & 1.764 & 0.557 \\[5pt]
\hline
Spin chain(chaotic) & 1.945 & 0.692 & 0.266 & 0.038 \\[5pt]
\hline
SYK model & 2.311 & 0.265 & 0.057 & 0.001 \\[5pt]
\hline
\end{tabular}

\begin{center}
\begin{tabular}{ |c|c|c|c|c| }
\hline
$\mathcal{I}_{T_3}(T)$ & 3-qubit & 4-qubit & 5-qubit & 8-qubit\\[5pt]
\hline
Spin chain(Integrable) & 7.589 & 4.240 & 2.894 & 0.329 \\[5pt]
\hline
Spin chain(chaotic) & 1.340 & 0 & 0 & 0 \\[5pt]
\hline
SYK model & 1.194 & 0 & 0 & 0 \\[5pt]
\hline
\end{tabular}
\end{center}
\caption{\label{Information backflow} The total amount of information backflow for different systems and different numbers of qubit. The top table considers $-I_3$, whereas the bottom one is for $-T_3$. Here, $T=40/g$ and $T=148/J$ for the spin-chains and the SYK-models, respectively. As a result, we can conclude that the system with a larger number of qubit tends to have smaller amount of information backflow.}
\end{table} 

In Fig.~\ref{finite}, we plot the numerical simulations of $-I_3$ and $-T_3$ for the integrable spin chain, chaotic spin chain, and the SYK model, involving different numbers of qubits. We can observe that as the qubit number decreases (increases), the oscillation magnitude of information scrambling for both integrable and chaotic dynamics increases (decreases). The result suggests that when the system size decreases (increases), it would be more likely (unlikely) to observe information backflow from non-local to local degrees of freedom.

Because any decrease of $-I_3$ ($-T_3$) signifies the backflow of information, we can quantify the total amount of information backflow within a time interval by summing up the total negative changes of the scrambling witnesses. More specifically, we define a quantity $\mathcal{I}_{Q}(T)$, which quantifies the total amount of information backflow for a given time interval $t\in[0,T]$, as follows
\begin{equation}
\mathcal{I}_{Q}(T)=\int_{t=0,\sigma_Q>0}^{t=T}\sigma_Q(t)~dt,
\end{equation}
where $Q \in \{I_3, T_3\}$ and $\sigma_Q(t) = \frac{d}{dt}Q(t)$. In other words, $\mathcal{I}_{Q}(T)$ integrates all positive changes of $Q$ (or equivalently, all negative changes of $-Q$) for $t\in[0,T]$. Note that this quantification of information backflow is consistent with that in the framework of quantum non-Markovianity~(see Ref.~\cite{chen2016quantifying}, for instance). We summarize the results in Table ~\ref{Information backflow}, which show that  as the number of qubit increases, the amount of information backflow $\mathcal{I}_{Q}(T)$ decreases, implying a stronger scrambling effect. 
%

\end{document}